\documentclass[preprint2,tighten,times]{aastex6}
\pdfoutput=1 
\usepackage{hyperref,verbatim}
\usepackage{amsmath,amstext,mathrsfs}
\usepackage[all]{hypcap} 
\newcommand*{\kh}{\color{black}}

\newcommand*{\HO}{\color{black} }

\begin{document}
\title{A comparison between Faraday Tomography and Synchrotron Polarization Gradients} 
\author{Ka Wai Ho\altaffilmark{1}, Ka Ho Yuen\altaffilmark{2}, Po Kin Leung\altaffilmark{1}, A. Lazarian\altaffilmark{2}}
\email{kyuen2@wisc.edu}
\altaffiltext{1}{Department of Physics, Chinese University of Hong Kong, Hong Kong}
\altaffiltext{2}{Department of Astronomy, University of Wisconsin-Madison, Madison, USA}
\begin{abstract}
Observations of synchrotron polarization at multiple frequencies in the presence of Faraday rotation can provide a way to reconstruct the 3D magnetic field distribution. In this paper we compare the well known Faraday Tomography (FT, \citealt{1966MNRAS.133...67B}) technique to the new approach named Synchrotron Polarization Gradients (SPG, \citealt{LY18b}). We compare the strengths and limitations of the two techniques, and describe their synergy. In particular, we show that in the situations when FT technique fails, e.g. due to insufficient frequency coverage, the SPG can still trace the 3D structure of magnetic field.
\end{abstract}



\section{Introduction}
\label{sec:intro}

Magnetic field structure is very important for key astrophysical processes in interstellar media (ISM) such as the formation of stars (see \citealt{MO07,MK04}), the propagation and acceleration of cosmic rays (see \citealt{J66,YL08}), the regulation of heat and mass transfer between different ISM phases (see \citealt{D09} for the list of the different ISM phases). Polarized radiation arising from the presence of the magnetic field is also important to explain the enigmatic CMB B-modes \citep{1997PhRvD..55.1830Z,2017PhRvL.118i1801C,2017MNRAS.464.3617K}. 

Synchrotron polarization is widely used in study magnetic fields structure in the sky. However, in the presence of the Faraday rotation it is not trivial to compensate for the distortion from the 2D polarization pattern within the volume emitting synchrotron radiation. The tracing of the actual three-dimensional (3D) magnetic field structure presents both a big attraction and an outstanding challenge. Potentially, by combining synchrotron data at different frequencies, one can try to obtain the magnetic field variation along the line-of-sight. \cite{1966MNRAS.133...67B} first suggested that the {\bf Faraday Tomography} (FT), i.e. multi-layer plane-of-sky magnetic field structures, can be obtained through proper Fourier transform from the polarized synchrotron emissions (See \citealt{2005A&A...441.1217B}, hereafter BB05). A number of works are coming out based on the depolarization of the synchrotron emissions \citep{2017MNRAS.468.2957R,2018arXiv181205399D,2018A&A...615..L3,2018IAUS..333..129H,2018MNRAS.474.3280F}.

A recently suggested alternative technique of magnetic field tracing employs the Synchrotron Polarization Gradients (SPGs, \citealt{LY18b}, LY18b). As discussed in the latter paper, the foundations of the SPGs are routed in the properties of MHD turbulence and turbulent reconnection \citealt{GS95,LV99}. As a result, the SPGs trace the {\it local} magnetic field through observationally resolved eddies. The applicability of the SPGs to the interstellar medium (ISM) arises form the fact that ISM is turbulent  \cite{AM95,CL09,CL10,2015ApJ...808...48B}. In the presence of Faraday Rotation, only a certain deepness of the synchrotron emission is {\it effectively} collected into the Stokes parameters. That means the synchrotron polarization map for a specific emitting frequency $f$ corresponds to the {\it plane-of-sky} magnetic field variation {\it accumulated} up to a certain depth along the line of sight.The theory of this effect in the presence of magnetic turbulence is given in \cite{LP16}. As a result, one can try to obtain the 3D magnetic field structure by utilizing multi-frequency synchrotron emissions. LY18b pointed out that by considering the differences of SPGs of polarized synchrotron maps obtained with multi-frequency observations, one can reconstruct the 3D magnetic field structure.

While the two proposals of tracing 3D magnetic field both rely on the multi-frequency synchrotron emission in the presence of Faraday Rotation, there are significant differences between the foundations of the two methods. One may wonder:  {\it (1) What are the limitations of the techniques? (2) How precise can the 3D B-field distributions can be traced with these techniques?  (3) Are the methods self-consistent?}. This paper is the first attempt to answer these important questions.

On one hand, the method of SPGs relies on the fact that turbulence is ubiquitous while the FT provides the self-consistent 3D mapping of the underlying regular magnetic field.  On the other hand, the method of FT has a much higher requirement on the number of frequencies compared to the SPG (e.g. Li et.al 2011b). In addition, the line-of-sight magnetic field strength information is not available for the FT but possible to obtain using SPG (see LY18b).  

We would therefore like to compare the two techniques in this paper through numerical simulations. Instead of confronting the techniques we search for their synergy.
In what follows, we briefly describe the numerical code and setup for simulation in \S \ref{sec:numerics}, the performance of two method in  \S \ref{sec:Result} \& \S \ref{sec:Tech}, the discussion of the synergy of two method in \S \ref{Sec:Diss} and summary in \S \ref{sec:Summary}.

\section{Numerical Simulations}
\label{sec:numerics}
\noindent {\bf Simulation setup.}
The numerical 3D MHD simulations were used already in \cite{LY18a} and LY18b by setting up a 3D, uniform, isothermal turbulent medium.  We use a range of Alfv\'{e}nic Mach number $M_A=V_L/V_A$ and sonic Mach number $M_s=V_L/V_s$, where $V_L$ is the injection velocity; $V_A$ and $V_s$ are the Alfv\'{e}n and sonic velocities respectively. The numerical parameters are listed in Table \ref{tt1} in sequence of ascending values of media magnetization $ \beta = 2 (M_A/M_S)^2 $.
\linebreak

\noindent {\bf Faraday Tomography.}
{\kh The concept of Faraday Tomography was first suggested by \cite{1966MNRAS.133...67B}. The method utilizes the fact that the Faraday rotation integral of synchrotron polarization along the line of sight is effectively a Fourier transform of the {\it complex polarized brightness per unit Faraday depth} $F(\phi)$ }:
\begin{equation}
\label{eq:FR1}
P(\lambda^2)= Q+iU =\int_{-\infty}^{\infty} F(\phi) e^{2i\phi\lambda^2} d\phi,
\end{equation}
{\kh where} $\lambda$ is the observed wavelength, $P(\lambda^2 )$ is the complex polarized surface brightness in terms of Stoke parameters $Q$ and $U$, and $\phi$ is the Faraday depth. {\kh Performing the inverse Fourier transform one can easily acquire $F(\phi)$}:
\begin{equation}
\label{eq:FR2}
 F(\phi)=\int_{-\infty}^{\infty} P(\lambda^2)e^{-2i\phi\lambda^2} d\lambda^2
\end{equation}
{\kh which provides the {\it 3D Magnetic field information as a function of $\phi$}. Since $\lambda^2$ only lies in the positive real space, the inverse Fourier transform cannot be computed accurately unless the negative part of $\lambda^2$ is provided. } BB05 provided a solution to the problem by introducing the window function $W(\lambda^2)$ to reconstruct $F(\phi) $. The window function is non-zero in the range of observed $\lambda^2$ and is otherwise zero. BB05 then defined the observed polarized surface brightness as:
\begin{equation}
\label{eq:BBP}
 \tilde{P}(\lambda^2) = P(\lambda^2)W(\lambda^2),
\end{equation}
{\kh As a result, the complex polarized brightness that includes the window $\tilde{F}(\phi)$ can be written as }
\begin{equation}
\label{eq:BBFR}
 \tilde{F}(\phi) = F(\phi) \ast R(\phi)=K\int_{-\infty}^{\infty} \tilde{P}(\lambda^2)e^{-2i\phi(\lambda^2-\lambda_0^2)} d\lambda^2,
\end{equation}
{\kh where $R(\phi)$ is the rotation measure transfer function (RMTF):}
\begin{equation}
\label{eq:RMTF}
R(\phi)=\frac{\int_{-\infty}^{\infty} W(\lambda^2)e^{-2i\phi(\lambda^2-\lambda_0^2)} d\lambda^2}{\int_{-\infty}^{\infty}W(\lambda^2)d\lambda^2} .
\end{equation}
The function $K$ is
\begin{equation}
\label{eq:WK}
K = \left(\int_{-\infty}^{\infty}W(\lambda^2)d\lambda^2\right)^{-1},
\end{equation}
and a parameter $\lambda_0$ is introduced to Eq. \eqref{eq:BBFR} \& \eqref{eq:RMTF}  in order to improve the the behavior of RMTF. The optimal $\lambda_0^2$ is the mean of $\lambda^2$ sample values obtained by the telescope. 
 
This technique introduced by BB05 is referred to as the Rotation Measure (RM) synthesis.  It shows promising in obtaining the 3D tomography magnetic field structure. The requirement for the technique to work is to have enough synchrotron polarization measurements at different frequencies. 

To calculate $F(\phi)$ in numerical simulations, we consider a column of data along {\kh the} line of sight (LOS) in {\kh a 3D MHD numerical data cube} and divide this column {\kh into $n$ segments}. Each segment contains the information from density, magnetic field (such as $Q$ \& $U$), and rotation measurement $\phi$. In this setting the polarization can be calculated as

\begin{equation}
\label{eq:Pcolumn}
P(\lambda^2) = \sum_{k=1}^{n} P_k e^{-2i\phi_k \lambda^2} ,
\end{equation}
where $P_k=Q_k+iU_k$, representing the $P_k$ at different $\phi_k$. The Faraday dispersion function can then be expressed as

\begin{equation}
\label{eq:Fcolumn}
F(\phi) = \sum_{k=1}^{n} \int_{-\infty}^{\infty} P_k e^{-2i\phi_k \lambda^2} e^{-2i\phi \lambda^2} d\lambda^2 .
\end{equation}

Although we cannot get the information of $P(\lambda^2)$ when $\lambda^2 <0$, it is still useful to assume  Eq. \ref{eq:Fcolumn} {\kh holds by assuming $P(\lambda^2<0)=0$} :
\begin{equation}
\label{eq:Fresult}
F(\phi) \approx \sum_{k=1}^{n} P_k \delta(\phi -\phi_k) ,
\end{equation}
{\kh which suggests that $|F(\phi)|$ can be decomposed into $n$ delta functions $\delta(\phi -\phi_k)$ peaked at $\phi_k$.} It is important to notice that Eq.~\ref{eq:Fresult} and the reconstructed Faraday dispersion function  $\tilde{F}(\phi)$ are not equivalent but share many similarities.
\linebreak
\begin{table}
 \centering
 \label{tab:simulationparameters}
 \begin{tabular}{c c c c}
Model & $M_s$ & $M_A$ & $\beta=2(\frac{M_A}{M_s})^2$\\ \hline \hline
Ms0.2Ma0.02 & 0.2 & 0.02 & 0.02 \\
Ms0.4Ma0.04 & 0.4 & 0.04 & 0.02 \\
Ms0.8Ma0.08 & 0.8 & 0.08 & 0.02 \\
Ms1.6Ma0.16 & 1.6 & 0.16 & 0.02 \\
Ms3.2Ma0.32 & 3.2 & 0.32 & 0.02 \\
Ms6.4Ma0.64 & 6.4 & 0.64 & 0.02 \\ \hline
Ms0.2Ma0.07 & 0.2 & 0.07 & 0.22 \\
Ms0.4Ma0.13 & 0.4 & 0.13 & 0.22\\
Ms0.8Ma0.26 & 0.8 & 0.26 & 0.22\\
Ms1.6Ma0.53 & 1.6 & 0.53 & 0.22\\\hline
Ms0.2Ma0.2 & 0.2 & 0.2 & 2 \\
Ms0.4Ma0.4 & 0.4 & 0.4 & 2 \\
Ms0.8Ma0.8 & 0.8 & 0.8 & 2 \\\hline
Ms0.13Ma0.4 & 0.13 & 0.4 & 18 \\
Ms0.20Ma0.66 & 0.20 & 0.66 & 18 \\
Ms0.26Ma0.8 & 0.26 & 0.8 & 18 \\\hline
Ms0.04Ma0.4 & 0.04 & 0.4 & 200 \\
Ms0.08Ma0.8 & 0.08 & 0.8 & 200 \\
Ms0.2Ma2.0 & 0.2 & 2.0 & 200\\\hline \hline
\label{tt1}
\end{tabular}
 \caption {Simulations used in our current work. The magnetic criticality $\Phi = 2 \pi G^{1/2} \rho L/B$ is set to be 2 for all simulation data. Resolution of them are all $480^3$.}
\end{table}

\noindent 
{\bf Synthesis of Position-Position-Frequency (PPF) cubes.}  {\kh We synthesize the PPF cubes following the procedures in  LY18b. We use the definition of synchrotron polarization in \cite{LP16} thus ignoring the wavelength dependences of synchrotron polarization arising from the cosmic ray spectrum. This means the source term $P_i ({\bf X}, z)$  will be wavelength-independent while the observed polarization will be wavelength-dependent due to Faraday rotation only. Similar to LY18b, we assume the cosmic ray index $\gamma=2$ since \cite{LP12} showed marginal effect of $\gamma$ on the spatial variations in the Stokes parameters. That means we can express the Stokes $Q$ and $U$ as: 
 \begin{equation}
 \begin{aligned}
 Q({\bf X}, z) &\propto p n_e (H_{x}^2(z)-H_{y}^2(z))\\
 U({\bf X}, z) &\propto p n_e 2H_x (z) H_y(z),
 \end{aligned}
 \end{equation}
}where $p$ is the polarization fraction, which is assumed to be constant, and $n_e$ is the density of relativistic electrons. The definitions of the Stokes parameters above correspond to the synchrotron intensity at the source $ I({\bf X}, z)\propto H_x^2(z) + H_y^2(z).$
\linebreak

\noindent
{\kh {\bf Effective measurable distance due to the Faraday screening effect.} In LP16 and LY18b the  {\it effective measurable distance} $L_{eff}$ arising from Faraday depolarization is introduced, refers to the line-of-sight distance over which the Faraday rotation phase of the synchrotron emission source is less than unity. Mathematically, 
}
\begin{equation}
\label{eq:el}
\frac{L_{eff}}{L} \sim \frac{1}{\lambda^2L} \frac{1}{\phi}
\end{equation}
{\kh where $L$ is the cloud thickness. Synchrotron polarization from distances larger than $L_{eff}$ contributed as noise in the resultant Stokes maps (see LY18b).}
\linebreak

\noindent 
{\bf Block averaging for gradient calculations}. Gradients of polarization are calculated by taking the values of polarization in the neighboring points and dividing them over the distances between the points following the recipe of \cite{YL17a}.  In this work, we focus on the smallest scale contribution as we did in \cite{LY18a}. {\kh LY18b provides the criteria for the gradients to be perpendicular to the magnetic field by investigating the indexes of the power spectrum and correlation function anisotropy. As we are using the same set of simulations used in LY18b, the criteria in LY18b are automatically satisfied. }
\linebreak\linebreak
\noindent {\bf Alignment Measure (AM)}. To quantify how good two vector fields are aligned, we employ the {\it alignment measure} that is introduced in analogy with the grain alignment studies (see \citealt{L07}):
\begin{equation}
AM=2\langle\cos^2\theta_r\rangle-1,
\end{equation}
(see \citealt{GCL17,YL17a}) with a range of $[-1,1]$ measuring the relative alignment between the {\it $90^o$-rotated} gradients and magnetic fields, where $\theta_r$ is the relative angle between the two vectors. A perfect alignment gives $AM=1$,  whereas random orientations generate $AM=0$. In what follows we use $AM$ to quantify the alignments of polarization gradients in respect to magnetic field.

\section{Results}
\label{sec:Result}
\subsection{Faraday tomography}
\subsubsection{The Reconstructed Faraday dispersion function of ISM}
\label{subsec:RFDF}

\begin{figure}
\centering
\includegraphics[width=0.49\textwidth]{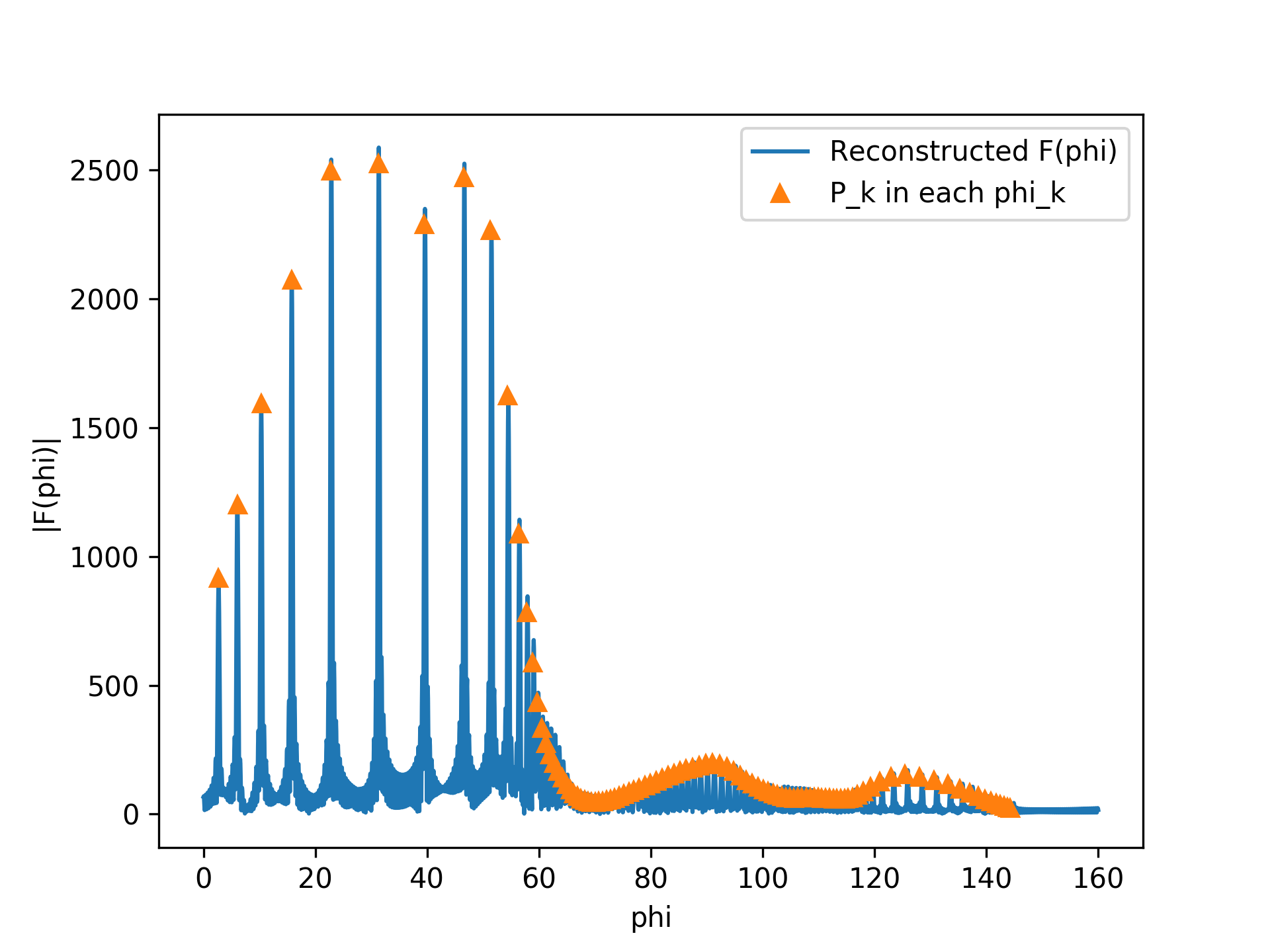}
\caption{\label{fig:Fsample} Comparison plot of between the actual $P_k$ and $\tilde{F}$. The samples are randomly chosen from a column of pixels along the line of sight in a 3D MHD simulation Ms3.2Ma0.32.}
\end{figure}

To use the Faraday tomography method, we use the reconstructed Faraday dispersion function $\tilde{F}(\phi)$ instead of Faraday dispersion function $F(\phi)$ because (i) as explained above, it is only possible to obtain $\tilde{F}(\phi)$ from observational data, (ii) both functions share similar spiky features with respect to the Faraday rotation measure, and (iii) the corresponding $\phi$-values of those peaks are similar for both functions. To demonstrate the points, Fig \ref{fig:Fsample} shows $\tilde{F}(\phi)$ from a simulation sample and compare with $F(\phi)$. They match the properties that Eq.~\ref{eq:Fresult} describes. The function $\tilde{F}(\phi)$ still contains the spiky feature we observed in $F(\phi)$. Moreover, the positions of the spikes in $\tilde{F}(\phi)$ are consistent to those in $F(\phi)$. Therefore, we can use Eq.\ref{eq:Fresult} and describe the features we calculated using the $\tilde{F}(\phi)$.

{\kh The reconstruction of Faraday dispersion function in simulation is rather trivial since the number of segments $n$ in Eq.\ref{eq:Fresult} corresponds to the number of frequency measurements in observations.  Potentially it should be possible to obtain the magnetic field distributions {\it exactly} provided that one has sufficiently large $n$ in Eq.\ref{eq:Fresult}. However, increasing the number of measurements is very costly as the Nyquist condition for the reconstruction of Faraday dispersion function grows with the square root of $n$; that means to increase the signal-to-noise by a factor of 2 one has to increase the number of measurements four times.  }

\begin{figure}
\includegraphics[width=0.49\textwidth]{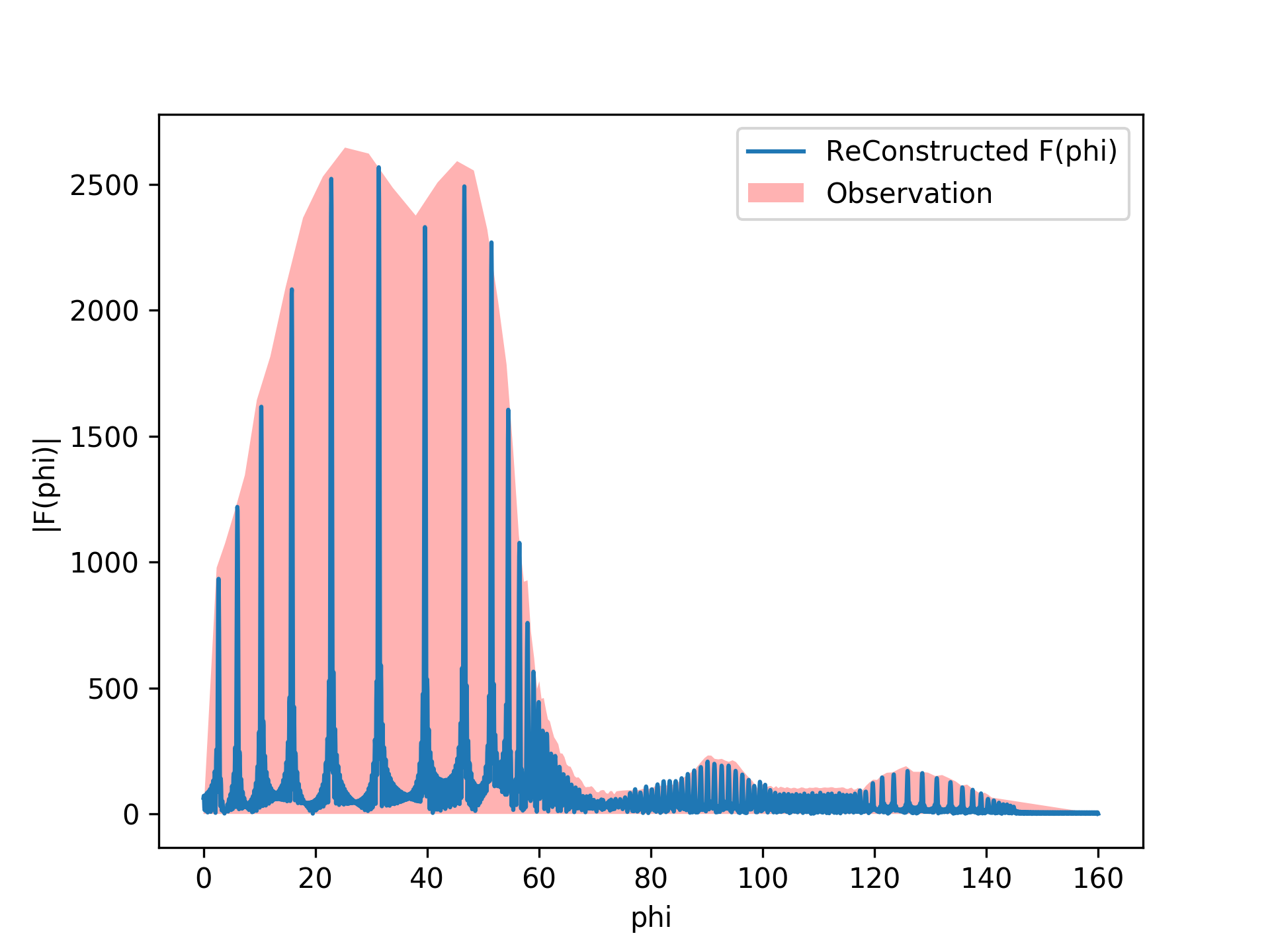}
\caption{\label{fig:FObservation} Illustration of the difference between simulation data and observation}
\end{figure}

{In Fig. \ref{fig:Fsample} we observed a lot of $\delta$-like structures sparsely spaced across $\phi$. We would refer the spaces between the $\delta$-like structures as ``gaps" while the $\delta$-like structures themselves are referred as "peaks". Fig. \ref{fig:FObservation} shows a simple illustration on how the number of ``peaks" would affect the detection accuracy of the Faraday tomography method. In observations (orange shaped region) one can only obtain a continuous distribution of the Faraday dispersion function in the Fourier space. However in numerical simulations the respective Faraday dispersion function is often composed of a number of discrete peaks, which correspond to different magnetic field values along the line of sight. The differences of the shape of the Faraday dispersion function between observations and simulations bring one very important question while applying the FT method in observation: How to determine the peaks in observation? 

To answer the aforementioned question, we performed a synthetic studies using numerical simulation and trying to mimic the observational settings. In observations the number of frequency channels $n$ is often limited, which makes the peak determination difficult. We therefore test whether the change of $n$ would result in different peak locations in the $F(\phi)-\phi$ plot. Fig.\ref{fig:Fsample} shows how the change of frequency channel resolution would alter the positions of peaks of the Faraday dispersion function. We see that positions of peaks are similar in both cases, showing that the resolution along the frequency axis does not affect locating the peaks of the Faraday dispersion function. }

\begin{figure}
\centering
\includegraphics[width=0.49\textwidth]{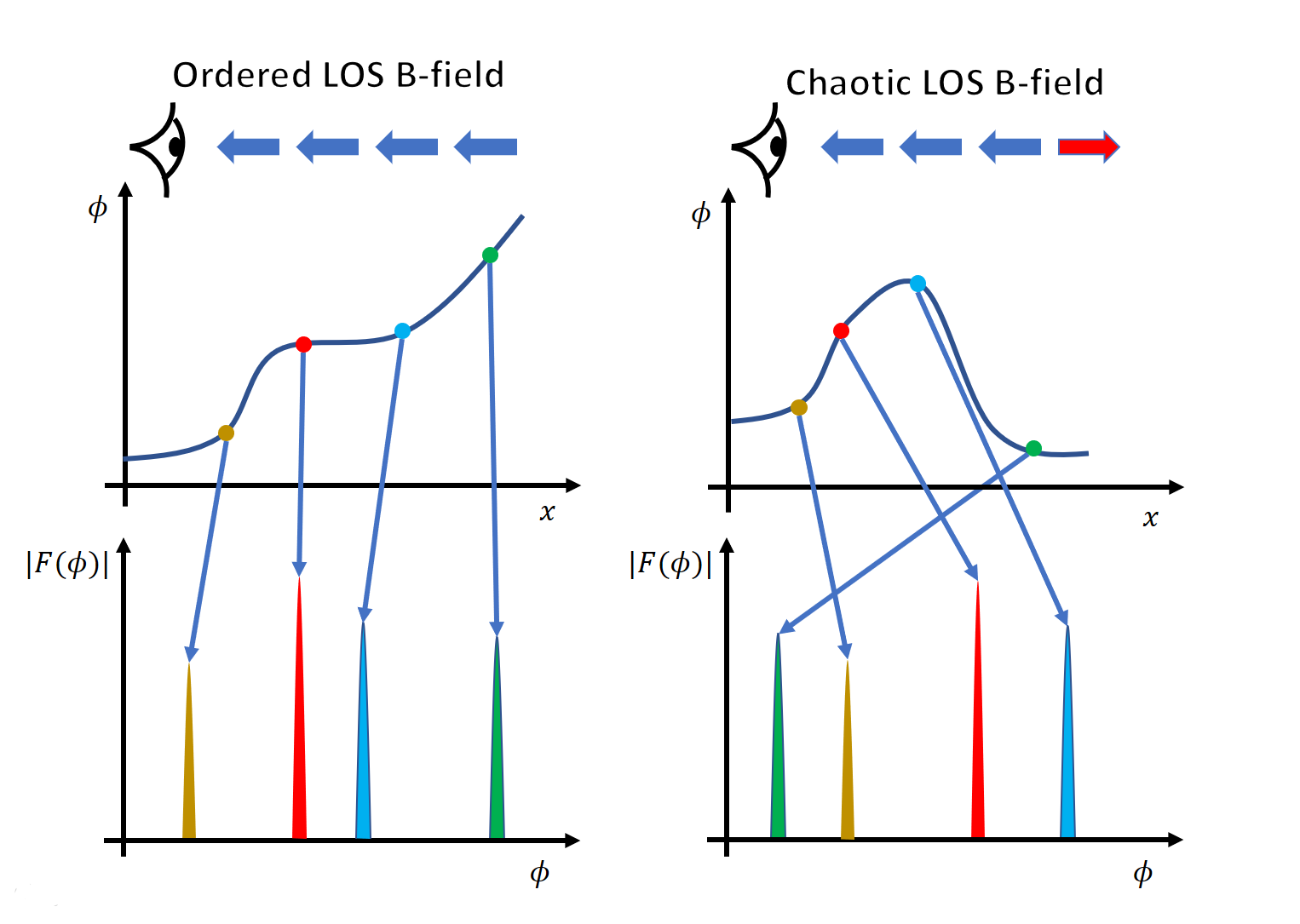}
\caption{\label{fig:Fillu} Illustration of two types of LOS B-field and how they affect the order of the $\delta$ function . }
\end{figure}

\subsubsection{\kh Determining the magnetic field orientation using Faraday Tomography}
{\kh 
The advantage of the Faraday Tomography over other methods is the high precision in determining the {\it plane-of-sky component} of the magnetic field as long as there are enough frequency channels that satisfy the Nyquist criterion. With a correct selection of frequency band in observation, one can determine the peaks in the Faraday dispersion function without difficulties. We would like to illustrate the power of Faraday Tomography using the numerical simulations listed in Table 1. 

Table \ref{tab:FTAM} shows the AM of the FT reconstructed {\it plane-of-sky component} magnetic field compared to the true magnetic field in numerical simulations. To get these results, we first convert the numerical cube to a Position-Position-Frequency (PPF) cube using the approach in \S \ref{sec:numerics} and then we compute the $\tilde{F}$.   After that, we locate the {\kh most significant} three {\it peaks for which the} real and imaginary part represent the Stokes parameters Q \& U.  The three peaks we located in the $\phi$ space are converted to the magnetic field measurements that we compare with the real magnetic field using the AM (See \S \ref{sec:numerics}). When we are computing the AM we randomly select 300 columns}\footnote{The whole process for FT is expensive since a two-step process is required to convert the simulation cubes to Faraday dispersion function, namely from the position-position-position (PPP) cubes to the PPF cubes and finally to PP$\phi$ cubes. Both processes are having the computational complexity of $O(N^3 \times N_f)$ and $O(N^3 \times N_{\phi})$, where N , $N_f$ and $N_{\phi}$ are the resolutions of simulations , number of frequency bands and number of rotation measure channels, respectively. We assume the $N_f=N_{\phi}$ in this study. Apart from that, to compute the AM, it also requires the computational complexity of $O(N^3 \times N_{\phi})$ to trace back the local maximum. So, checking pixels randomly provides a more efficient approach in checking the accuracy of FT.} We also use the $\phi$ value found from those three peaks and compare to the the exact value of the polarization angle available from the simulations. Finally, we compare both angles and get the AM values for each of them. There are subtle differences between the cases whether the mean field or the turbulent field dominates along the line of sight \citep{LP16} and careful studies have to be done these cases. To test this, we perform tests in both cases with ordered and chaotic LOS B-fields. For the ordered B-field case, we rotate the simulation cube such that the mean field direction is pointing to the observer. For the chaotic field case, we further rotate the cube so that the mean field direction would be parallel to plane of sky. One can see that for both cases of ordered and chaotic LOS B-field the AM is pretty high, with an average value of $\sim 0.7$ - $0.8$, if we {\kh pick} the correct frequency band.

\begin{table}
\centering
\begin{tabular}{c c c c}
LOS B-field type: & Ordered & Chaotic & Chaotic \\ \hline \hline
Frequency (Hz): & 3x$10^{8}$-3x$10^{11}$ & 3x$10^{8}$-3x$10^{11}$ & 3x$10^{7}$-3x$10^{11}$ \\ \hline \hline
Model & AM  & AM  &  AM \\ \hline \hline
Ms0.4Ma0.04 & 0.92 & 0.24 & \textbf{0.70} \\
Ms0.8Ma0.08 & 0.90 & 0.26 & \textbf{0.81} \\
Ms1.6Ma0.16 & 0.88 & 0.28 & \textbf{0.81} \\
Ms3.2Ma0.32 & 0.87 & 0.26 & \textbf{0.74} \\
Ms6.4Ma0.64 & 0.86 & 0.23 & \textbf{0.60} \\
Ms0.4Ma0.32 & 0.82 & \textbf{0.99} & 0.27 \\
Ms0.8Ma0.264 & 0.81 & \textbf{0.95} & 0.28 \\
Ms1.6Ma0.528 & 0.76 & \textbf{0.70} & 0.23 \\
Ms0.4Ma0.4 & 0.83 & \textbf{0.95} & 0.87 \\
Ms0.8,Ma0.8 & 0.65 & \textbf{0.73} & 0.40 \\
Ms0.132Ma0.4 & 0.78 & \textbf{0.94} & 0.94 \\
Ms0.264Ma0.8 & 0.63 & 0.55 & \textbf{0.73} \\
Ms0.04,Ma0.4 & 0.74 & 0.92 & \textbf{0.94} \\
Ms0.08,Ma0.8 & 0.75 & \textbf{0.72} & 0.64 \\
 \hline \hline
\end{tabular}
\caption{\label{tab:FTAM} Accuracy of FT method for different frequency bands and LOS B-field. The bold font is use to emphasize the values of AM corresponding to high alignment.}
\end{table}

\subsubsection{Possible improvement of Faraday Tomography and its impact}

It is worth mentioning that {\kh we do not apply the technique involving a multiplicative factor of $ e^{2i\phi\lambda_0^2}$ } in the calculation of the $\tilde{F}$ as suggested in BB05. {\HO BB05 thought that this factor could influence the phase rotation of $\tilde{F}$ in both real and imaginary space thus affecting the tracing power of FT. However, we have shown that FT performs well even without this factor.} Also, with the extra factor, Eq.\ref{eq:Flamaba0} will become:
\begin{equation}
\begin{aligned}
\label{eq:Flamaba0}
F(\phi) = \sum_{k=1}^{n} \int_{-\infty}^{\infty} P_k e^{-2i\phi_k \lambda^2} e^{-2i\phi (\lambda^2-\lambda_0^2)} d\lambda^2\\
\approx \sum_{k=1}^{n} P_k e^{2i\phi_k\lambda_0^2}\delta(\phi -\phi_k)
\end{aligned}
\end{equation}
Comparing to Eq.\ref{eq:Fresult},  Eq.\ref{eq:Flamaba0} contains an extra term $e^{2i\phi_k\lambda_0^2}$ . This term will not affect the features we explained about the $|F(\phi)|$ at $\phi$ space since $e^{2i\phi_k\lambda_0^2}$ will be canceled during the calculation by its conjugate term. So, the peak value and the location of the $\delta$ function will not be changed. However, if we use the same treatment as we did on $|F(\phi)|$ to  $Q_k$ \& $U_k$, then the extra  term $e^{2i\phi_k\lambda_0^2}$ will not be cancelled but induce extra Faraday rotation. In fact, there is a similar mathematical origin for the expressions from Faraday rotation $e^{-2i\phi_k\lambda^2}$ and the term in Eq.\ref{eq:Flamaba0} that is $e^{2i\phi_k\lambda_0^2}$. As a result, the inclusion of the $e^{2i\phi_k\lambda_0^2}$ term will introduce an additional rotation of $-\phi_k\lambda_0^2$ degree for the polarization angle $\phi_k$. The removal of the $e^{2i\phi_k\lambda_0^2}$ term would make the polarization angle more physically justified.

\subsection{The performance of SPG in tracing 3D magnetic field}
\label{subsec:SPGResult}
To compare the performance of the SPG tomography proposed in LY18b with the FT , we divide the axis along LOS into 20 slides, which corresponds to $L_{eff}/L$ as from 0.05 to 1.0 with a separation of 0.05. This allows us to trace the 2D magnetic field structure at different depth. The {\kh frequencies} required are computed from the Eq.\ref{eq:el}. 

It is important to note that the rule to choose the frequency band is different for SPG and FT. Suppose we get two measurements from $P(f_{i})$ and $P(f_{i+1})$ at any two neighbouring frequency points $f_i$ and $f_{i+1}$, with the definition that the frequency width is defined as the difference of frequencies $\Delta f = f_i-f_{i+1}$. In the case of SPG, since the relation $L_{eff} \sim \frac{f^2}{\phi}$ from the Eq.\ref{eq:el} holds for all frequencies, the {\it effective line of sight thickness} $\delta L_{eff}$ is also related to the frequency width following Eq.\ref{eq:el} that $\delta L_{eff} \sim \delta f_{0}^2 = f_{i} ^2 - f_{i+1} ^2 $. If one wants $\delta L_{eff}$ to be fixed, then the differences of frequency bands follows quadratically with the above relation. In the case of FT, since we are performing Fourier transform with the term $e^{2i\phi \lambda^2}$ in $\lambda^2$ space, the selection of frequency bands would then depend on the fact that the differences of wavelengths have to be constant $ \lambda_2 ^2 - \lambda_1^2 = \delta \lambda_{0}^2$. The dependencies of frequency bands as a result is inversely quadratically. We would discuss more in \S \ref{subsec:Frequencysampling} .

After a correct selection of frequency bands, which corresponds to a set of effective line of sight thicknesses, we can then get the synchrotron polarization derivative map from calculating the difference of polarized intensity $\bar{\Phi}=\sqrt{\Delta Q^2 + \Delta U^2}$ by:
\begin{equation}
\begin{aligned}
\label{eq:SPGQU}
\Delta Q =Q(f_{i+1})-Q(f_{i}),\\
\Delta U =U(f_{i+1})-U(f_{i}),
\end{aligned}
\end{equation}
where $f_{i+1}$ and $f_{i}$ are two neighboring frequencies. The two maps contain the information of cumulative magnetic morphology in the corresponding depth and computing the differences of gradient orientation would determine the 3D magnetic morphology between the two line of sight depths $\delta L_{eff} = L_{eff}(f_{i+1}) - L_{eff}(f_{i})$. We use the block averaging technique (see YL17) to obtain the statistical measurement of gradient orientation within a sampling region. In our calculations we choose the block size to be $30\times30$ pixels. Table \ref{tab:SPGAM} shows the AM from different numerical cubes following our treatment.

\begin{table}
\centering
\label{tab:SPGAM}
\begin{tabular}{c c c c}
LOS B-field type: & Ordered & Chaotic  \\ \hline \hline
Model & AM  & AM  \\ \hline \hline
Ms0.4Ma0.04 & 0.21 & 0.50 \\
Ms0.8Ma0.08 & 0.25 & 0.63 \\
Ms1.6Ma0.16 & 0.39 & 0.70 \\
Ms3.2Ma0.32 & 0.23 & 0.68 \\
Ms6.4Ma0.64 & 0.25 & 0.50 \\
Ms0.4Ma0.32 & 0.17 & /    \\
Ms0.8Ma0.264 & 0.19 & 0.48 \\
Ms1.6Ma0.528 & 0.26 & 0.50 \\
Ms0.4Ma0.4 & /    & /    \\
Ms0.8Ma0.8 & /    & /    \\
Ms0.132Ma0.4 & /    & /    \\
Ms0.264Ma0.8 & /    & /    \\
Ms0.04Ma0.4 & /    & /    \\
Ms0.08Ma0.8 & /    & /    \\
 \hline \hline
\end{tabular}
\caption{Accuracy of SPG method in difference LOS B-field,The columns without AM values correspond to the case that the required frequency to calculate the SPG is less then $10^7Mhz$.}
\end{table}

A clear trend seen from the table is that SPG traces the magnetic morphology with higher accuracy compared to that for the chaotic LOS field. This follows from the differences in the localization of $L_{eff}$ for the case of regular and chaotic field \citep{LP16}. For the ordered LOS magnetic field case we get the AM around  0.2 to 0.3 which agrees well with the results in LY18b. All gradient techniques share the same foundation of anisotropic MHD turbulence, i.e. the turbulent eddies are elongated along local magnetic field directions which were described in \cite{GS95} and \cite{LV99}. The property of anisotropy will affect the structure of other observables, e.g. integrated intensities, velocity centroids, velocity channels and also synchrotron intensities. The performance of gradient techniques are highly related to whether the environment is dominated by turbulence or not, and how anisotropic the system it is. In our case of ordered B-field, the mean magnetic field direction is pointing along the LOS. Since the turbulent eddies are all aligned to the local field and most of our numerical cubes are having strong magnetization, when observing along the mean field only a weak or even no anisotropy can be detected. This also explains why the model with lower $M_A$ traces magnetic field better.

\section{\kh What techniques should we select?}
\label{sec:Tech}
\subsection{LOS information {\kh along} the $\phi$ axis}
\label{subsec:RFDFX}

In order to know the information of magnetic field from different depths of $\tilde{F}(\phi)$, we should first study the relation between the $\delta$ functions at $\phi$ and LOS axes. On the scale of the resolution, {\kh one can see} that the correct identification of the position of $\delta$ function is closely related to how ordered is the LOS B-field. {\kh Depending on the strength of the mean and fluctuating magnetic field, we can classify the magnetic field conditions into two cases.} If the LOS B-field is highly ordered, e.g. when the B-field direction is pointing either towards or away from the LOS observer, each $\phi_k$ is unique and we can relate the order in the location of the $\delta$ function in the reconstructed distribution and the order in the location of emitters in the source. {\kh If} the LOS B-field is chaotic, i.e. is pointing both towards and away from observer, this relation fails.  Figure \ref{fig:Fillu} illustrates both  cases. In a magnetized environment, the LOS magnetic field structure is usually determined by the orientation of the mean field in respect to the observer. For the SPGs problems arise when the mean field is directed exactly either towards or away form the observer. o solve the problem that there is a complicated dependence of $\phi$ to LOS, a physical model of local interstellar medium should be built up to relate the two physical quantities \citep{2015A&A...583..A137,2017A&A...597A..98V}. However, it requires other measurements for building it up. 

\subsection{Synergy of SPGs \& FT}
\label{subsec:synergy}
The SPGs is a very new technique and therefore the corresponding procedures of restoring the 3D structure do require refinement and improvements. We expect that the AM in Table 3 will improve as the technique is getting mature. Nevertheless, even at this point we can clearly see the synergy of the SPGs and the FT.

As we showed in section \ref{subsec:RFDF}, the LOS B-field structure is critical for determining the positions of synchrotron emission sources in the $\tilde{F(\phi)}$. Since the LOS B-field is turbulent in the ISM, tracing 3D B-field structure with FT is difficult in many cases, especially if the mean field is close to perpendicular to the line of sight. This is exactly the case, however, when the SPGs can be most useful. As we discussed above the SPGs have difficulty with tracing magnetic field structure when the mean field is nearly parallel to the line of sight. At the same time, SPG performs the best way when the mean magnetic field is perpendicular to the line of sight.  

\begin{figure}
\centering
\includegraphics[width=0.49\textwidth]{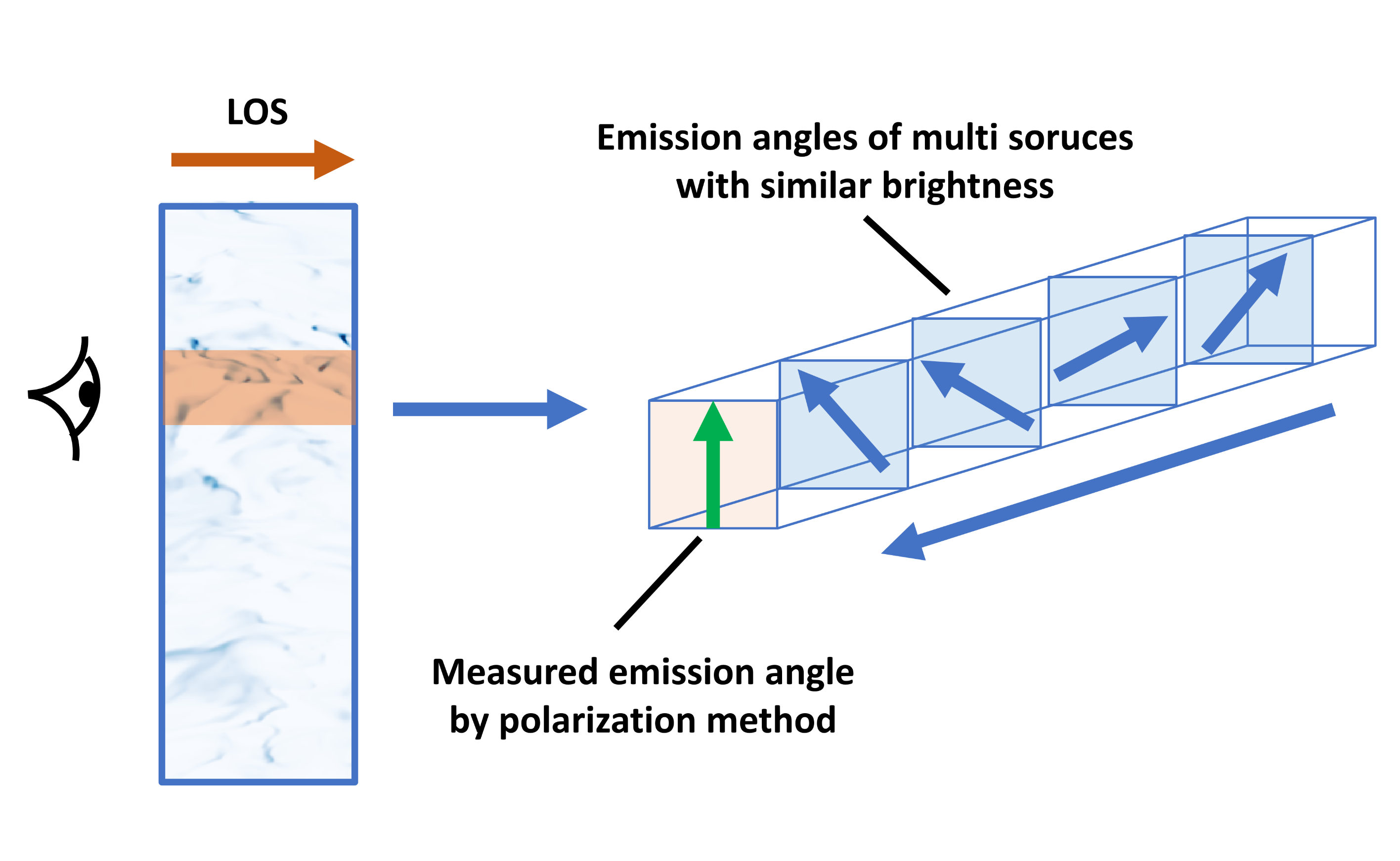}
\caption{\label{fig:MultiSource} Illustration of how multi emission source with similar intensity change the result of the emission angle measurement  .}
\end{figure}

The traditional polarization method {\kh that used widely in the astronomical community} provides one polarization angle vector {\kh per} pixel. {\kh It is usually being interpreted to be} the 2D B-field direction at that pixel. However , this polarization direction that we measured is actually a sum of the Stokes parameters of that column and the direction of polarization is affected by the Faraday rotation (see Eq.\ref{eq:Pcolumn}). While the Faraday rotation effect can be minimized by observing the polarization at  high frequencies, even in this case the angle that we measure is a mean angle of synchrotron emission from multiple synchrotron sources along the LOS. If the LOS emission is dominated by one bright emission source, the measured polarization angle is close to the polarization angle at the source. In an extreme case where we have multiple sources with comparable brightness, the measured polarization angle may be misleading. As a simple example let us consider two synchrotron polarization sources of similar brightness having 0 and $\frac{\pi}{4}$ polarization angle, which is illustrated by Fig \ref{fig:MultiSource}. The angle measured in this case is $\sim$ $\frac{\pi}{8}$ and this fails to represent the underlying magnetic field information for either of the source. The situation of having multiple bright sources along the line of sight is very common. Unfortunately, it is impossible to tell how many sources are there along the line of sight with their respective weight by simply increasing the observing frequency.

On the contrary, the FT method (\citealt{1966MNRAS.133...67B}, BB05) has ability to detect the number of the emission sources along the LOS. The amplitude of $\delta(\phi)$ functions at the Faraday dispersion function representing the brightness of each source. By counting the number of {\kh peaks of the Faraday dispersion function}, not only we know the numbers of intensive sources along the LOS, also the 2D magnetic field structure within the source. 

The mathematical nature of the FT method provides the ability to depolarize emission along the LOS. We can even know the intensity distribution along LOS from this method. It also {\kh gives a} high AM to trace the magnetic field orientation whichever direction the mean field points to. However, the result from FT was limited in the scale of pixel that explained in Section \ref{subsec:RFDF}. We lose the positional information during the process from $\lambda^2$ to $\phi$ space. We cannot get any information about the magnetic field morphology of different depth even in the local scale. Nevertheless, we cannot construct the 3D field information due to the lack of the LOS B-field information. In this sense, the FT method acts like a probe that allows one to obtain precise 2D field information for each source but fails to provide any of the positional nor line-of-sight B-field information.
  
On the other hand, SPG {\kh depicts} the magnetic field structure quite well. It can map the 2D magnetic morphology in {\kh different} depths and even the 3D structure shown in LY18b. Unlike FT which is applied to all the environment, this method {\kh comes} with requirements of the environment. {\kh In principle the method of SPG requires the system to have anisotropic turbulence satisfying the spectral conditions suggested by LY18b, namely the spectral slope has to be steeper than $-1$ and the local anisotropy has to be along local magnetic field direction. In fact, the electron spectrum studies from \cite{AM95} and later extended by \cite{CL10} showed that turbulence spectrum is $-5/3$ for 15 orders of magnitudes, including the scale where synchrotron emissions are significant. Moreover, the anisotropy of the statistical measures available through observations is a well established fact proven with both synthetic observations and actual observational data (see \citealt{Letal02,EL05, 2008ApJ...680..420H,brazil18}).\footnote{The anisotropy of MHD turbulence has been has been known for a while (see \citealt{1984ApJ...285..109H}). It is also a part of \cite{GS95} picture. It is essential for the SPGs, however, that the anisotropy is present not in terms of mean magnetic field, but in terms of {\it local} magnetic field, i.e. the magnetic field at the location of turbulent eddies. This concept is not a part of the original \cite{GS95} model, but it was introduced in later publications. It follows from turbulent reconnection theory \citep{LV99} and is supported by numerical simulations \cite{CV00,2001ApJ...554.1175M, CLV02}.}}  

As a result, even though both techniques advertised themselves that they can provide magnetic field information tomographically, their products are different and it is difficult to have direct comparisons between them. For instance, one can identify the bright sources using FT while constructing the 3D field morphology using SPG. The information acquired by both methods are complementary but a side-by-side comparison requires further conversion between the two methods.

\begin{table}
\centering
\label{tab:SPGFTtable}
\begin{tabular}{c c c}
  & SPG & FT  \\ \hline \hline
2D/3D Structure & Yes & No \\
Order Info  & Yes & Generally No \\
object obtained & local feature & Intensity at $\phi$ \\
Distribution of Intensity & No & Yes \\
Data Pt. Requirement & Less & More \\
Accuracy & Relatively Low & Depend to Freq. Den. \\
 \hline \hline
\end{tabular}
\caption{Table to summarize the difference between SPG and FT method }
\end{table}

\subsection{Frequency sampling between the SPG and FT}
\label{subsec:Frequencysampling}

As we discussed in \S \ref{subsec:SPGResult}, there is a crucial difference in choosing the width of the frequency band for the two techniques. The dependencies of the frequency width in FT is related to $\delta \lambda^2$ and SPG to $\delta \frac{1}{\lambda^2}$. Not only that, they also have different meanings when changing the frequency width. For FT, BB05 brings out the following relation between $\phi$ and $\delta\lambda^2$, as 
 \begin{equation}
\begin{aligned}
\label{eq:FTdeltalamda}
||\phi_{max}|| \approx \frac{\sqrt{3}}{\delta\lambda^2}.
\end{aligned}
\end{equation}
The $\delta\lambda^2$ term controls the maximum $\phi$ that is not affected by the $m\pi$ ambiguities problem. To maximize the usable range of $\phi$, the  robust way is to narrow the frequency width. Sometimes it is necessary for observer to narrow the width if the synchrotron emission along LOS is lining on a large $|\phi|$. Therefore, the FT requires more data points to keep its accuracy.   

For the SPG technique, we can get the relation from  Eq. \ref{eq:el} and bring out the result 
\begin{equation}
\begin{aligned}
\label{eq:SPGdeltalamda}
\delta L_{eff} \propto \delta \frac{1}{\lambda^2}.
\end{aligned}
\end{equation}
The width of the frequency band controls how thick is the layer along LOS. Increasing the number of data points within the frequency band can get us a more detailed morphology within the region.  One of the advantages of this method is that the result will not be affected by $m\pi$ ambiguities problem. Observers have the flexibility to choose the frequency band width.

\section{Discussion}
\label{Sec:Diss}
\subsection{Requirements for the instruments}

It has been shown in \S \ref{subsec:synergy} that SPG and FT techniques are actually complementary to each other.  The SPG method requires far less data points compared to FT, and it can be used to scan though the whole region. The method can then provide a 2D magnetic field structure of the whole environment with good accuracy  (AM~0.6-0.7). Apart from tracing the direction of magnetic field, the SPG can be used to test the magnetization information. The corresponding technique of obtaining a distribution of Alfvenic Mach numbers $M_A$ was demonstrated in LY18a.

\begin{figure*}
\centering
\includegraphics[width=0.98\textwidth]{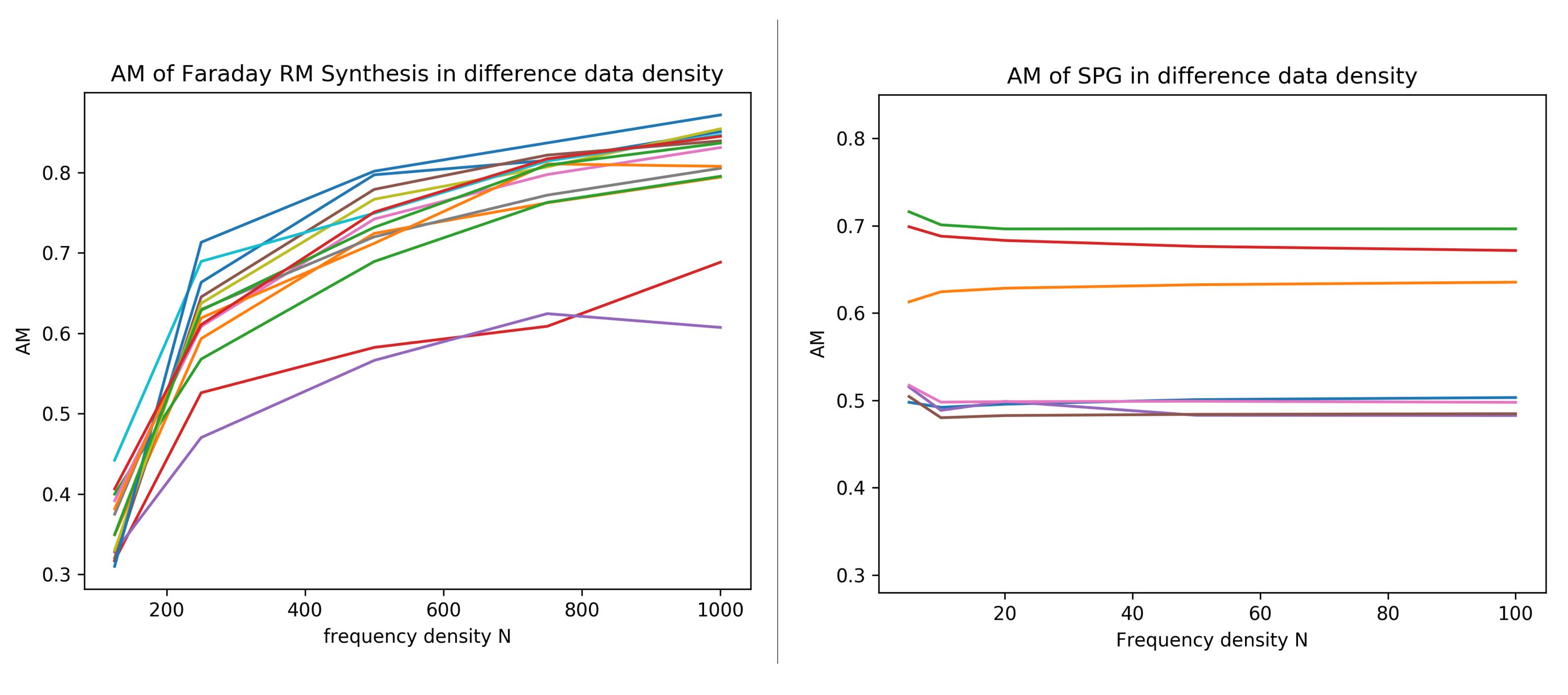}
\caption{\label{fig:AMFreqdensity} The plot of the AM (y-axis) and frequency density(x-axis) in difference numerical simulation listed in section 2. Every color line represents a numerical simulation. The left panel is represent the plot of FT and right panel is SPG. }
\end{figure*}

FT technique (\citealt{1966MNRAS.133...67B}, BB05, see also \citealt{2018arXiv181205399D,2018A&A...615..L3,2018IAUS..333..129H,2018MNRAS.474.3280F}) provides a high accuracy restoration of magnetic field (AM~0.8-0.9) in ideal case. However, it requires more frequency measurements to keep the high resolution. From BB05, It brings out the resolution of the Faraday dispersion function as,
\begin{equation}
\begin{aligned}
\label{eq:FDFResolution}
\delta \phi = \frac{2\sqrt{3}}{\Delta \lambda^2}.
\end{aligned}
\end{equation}
One can easily see that the resolution of the Faraday dispersion function is very sensitive to the frequency density by combining Eq. \ref{eq:FTdeltalamda} and Eq. \ref{eq:FDFResolution}. 
\begin{equation}
\begin{aligned}
\label{eq:FDFResolution2}
\frac{\delta \phi }{\phi_{max}} = \frac{2\delta \lambda^2}{\Delta \lambda^2} = \frac{2}{N}.
\end{aligned}
\end{equation}
where $N=\frac{\Delta \lambda^2}{\delta \lambda^2}$ is the {\it frequency density} in a certain frequency band, meaning the total number of N measurements observed in a particular frequency band $\Delta \lambda^2$ with frequency width $\delta \lambda^2$. The resolution would then affect the accuracy of the FT technique to probe the direction of the magnetic field. To test how the frequency density affects the accuracy, we perform a test for both techniques with different frequency density in Fig.\ref{fig:AMFreqdensity}.

There is a clear trend that the accuracy of the FT technique would drop with respect to the frequency density in all of our simulations. For instance, when the frequency density drops to 250, the accuracy of FT drops to the range of $0.5-0.7$. It is worth noticing that the SPG technique can achieve the same accuracy with only 20 frequency points. The performance of FT becomes worse when one further decreases the frequency density to 125 frequency points. On the other hand, SPG shows a very stable result with only small fluctuation even in the very low-frequency density measurement like 5 and 10 frequency points. 

We have demonstrated in \S \ref{subsec:SPGResult} that SPG is flexible when applying to data with different frequency resolution and plausible in constructing 3D magnetic field morphology with high precision. To better resolve the Faraday depth structure, FT requires high precision instrument like The Low-Frequency Array(LOFAR). In view of this, it is more favorable to use the SPG in tracing the 3D magnetic field due to the low-frequency band and the limited frequency resolution we have in current observation since the resolution is comparable to the maximum detectable scale in FT.\citep{2015A&A...583..A137,2018ApJJohn}

\subsection{The future of FT and SPGs}
As a member of VGT family, SPG also based on the same property of anisotropic MHD turbulence. Recently VGT is getting more mature especially in the case of extracting field information from molecular clouds. For instance, a new technique like Moving Window Method was developed to correct the gradient direction and improve the alignment between the gradient direction and magnetic field direction \citep{LY18a}.  In addition, the extraction of magnetization using the dispersion of velocity gradients in diffuse HI media are developed recently \citep{LYH18}. Those methods were already applyed to both diffuse interstellar media and  self-gravitating molecular clouds and provide reliable result \citep{HuNatureSub}. Likewise, those techniques applied to VGT could also migrate to SPG in the future. By obtaining the multi-frequency synchrotron emission, SPG could construct a 3D magnetization distribution of the interstellar media, by analyzing the gradient distribution in different effective depth of SPG. The accuracy of such predictions can be improved through the Moving Window Method. In the near future, SPGs could provide not only the high precision 3D magnetic field morphology but more physical information along the line of sight. 

The use of FT is also combined with other recently developed technique to provide further analysis of the observation data, e.g., combining the Rolling Hough Transform (RHT, \citealt{2015PhRvL.115x1302C}) with FT \citep{2018A&A...615..L3}. 

In the paper, they are using RHT to characterize the properties of the straight depolarization canals from the LOFAR observation data. As the result, they provide the relative orientation analysis of the Faraday dispersion Map. Moreover, it is worth mentioning that the similar analysis of relative orientation can be obtained also by SPG with a much higher accuracy and a possible 3D morphological structure based on the data provided. We believe that the aforementioned filamentary pattern/structure are the result of velocity crowding and related to the velocity gradient. In the future, SPG itself could also be combining with FT in observation data to obtain the orientation analysis.


\section{Summary}
\label{sec:Summary}
\label{sec:conclusion}
The present paper compares the two techniques named Faraday Tomography and Synchrotron Polarization Gradients that are plausible to trace the 3D magnetic field using polarized synchrotron emission. We have explored numerically the performance of these techniques and analyzed their strengths and limitations. The FT method can provide high accuracy in magnetic field tracing provided that a high frequency coverage is available and the mean magnetic field are not close to perpendicular to the line of sight. It is advantageous that the SPGs, on the other hand, provide a better accuracy in tracing the 3D magnetic field when the magnetic field is nearly perpendicular to the line of sight. Moreover, we demonstrated that the SPGs require less frequencies when restoring the magnetic field structure. As a result, we claim that combining the two techniques for acquiring the 3D structure of interstellar magnetic field is advantageous, especially as the SPG technique matures and gets more accurate.

\section*{Acknowledgements}
AL acknowledges the support of NSF grants DMS 1622353, AST 1715754 and 1816234.







\end{document}